%% file: main.tex
\documentclass[a4paper]{article}

\usepackage{amsmath,graphicx}
\usepackage{amssymb}
\usepackage{hyperref}
\usepackage{cite}
\usepackage{url}
\usepackage{bm}
\usepackage{color}
\usepackage{mathtools}
\usepackage{newtxtext,newtxmath}
\usepackage{INTERSPEECH2021}

\title{Parallel Tacotron~2: A Non-Autoregressive Neural TTS Model \\ with Differentiable Duration Modeling} 
\name{Isaac~Elias$^1$,~~Heiga~Zen$^2$,~~Jonathan~Shen$^3$,~~Yu~Zhang$^3$,~~Ye~Jia$^3$,~~RJ~Skerry-Ryan$^3$,~~Yonghui~Wu$^3$}
\address{
    $^1$Google, Israel\\
    $^2$Google, Japan\\
    $^3$Google, USA
}
\email{isaace@google.com, heigazen@google.com}

\begin{document}

\input{mymacro.tex}
\maketitle
\begin{abstract}
This paper introduces \emph{Parallel Tacotron~2}, a non-autoregressive neural text-to-speech model with a fully differentiable duration model which does not require supervised duration signals. 
The duration model is based on a novel attention mechanism and an iterative reconstruction loss based on Soft Dynamic Time Warping, this model can learn token-frame alignments as well as token durations automatically.
Experimental results show that Parallel Tacotron~2 outperforms baselines in subjective naturalness in several diverse multi speaker evaluations.
\end{abstract}

\noindent Published in INTERSPEECH 2021

\section{Introduction}
The neural text-to-speech (TTS) approach has made significant impact on research and development for the last five years \cite{oord-arxiv-2016,char2wav,tacotron,oord2018parallel,shen-arxiv-2020,li-aaai-2019,ren2020fastspeech2}.  
Tacotron~2 \cite{tacotron2} is one of the popular neural TTS models in the research community.
It combines an encoder-decoder model using Soft Attention \cite{bahdanau2014neural} and predict Mel-spectrogram given characters. This is combined with a neural audio generation model \cite{oord-arxiv-2016,kalchbrenner2018efficient} which generates the waveform given the predicted Mel-spectrogram.
Although Tacotron~2 can synthesize naturally sounding speech, there are two shortcomings.
(1) The use of the Soft Attention can introduce robustness errors, such as over-generation (\textit{e.g.}, word repetitions) and under-generation (\textit{e.g.}, word skipping) \cite{he2019robust,ForwardBackwardAttention,guo-interspeech-2019,shen-arxiv-2020}.
(2) Due to the use of a recurrent neural networks (RNN) in both the encoder and decoder, both training and inference are not efficiently executed on modern parallel accelerators; such as graphics processing units (GPUs) or tensor processing units (TPUs).

\emph{Parallel Tacotron} \cite{ptaco} is a non-autoregressive neural TTS model augmented by a variational auto-encoder (VAE)-based residual encoder \cite{hsu2018hierarchical,sun2020fullyhierarchical}.
Like other duration-based non-autoregressive neural TTS models \cite{ren2019fastspeech,ren2020fastspeech2,zeng-icassp-2020,donahue-arxiv-2020,miao2020efficienttts,lim-arxiv-2020}, this architecture can address these shortcomings.
As it based on token durations, it is less prone to synthesize speech with robustness errors and at the same time it is easier to control rhythm by modifying the predicted token durations.
However, it relies on an external aligner that provide supervised duration signals.
This requirement increases the complexity of its training process and makes the model sensitive to the performance of an external aligner.
To address this dependency, this paper introduces \emph{Parallel Tacotron~2}, which is an extension of Parallel Tacotron.\footnote{Audio examples: {\scriptsize \url{https://google.github.io/tacotron/publications/parallel_tacotron_2/index.html}}.}
It includes (1) a fully differentiable duration model, (2) a learned upsampling mechanism using attention with a novel auxiliary context, and (3) an iterative reconstruction loss based on Soft Dynamic Time Warping (Soft-DTW)  \cite{softdtw}.
As such, Parallel Tacotron~2 requires neither supervised duration signals nor teacher forcing of target durations at training time. 
The fully differentiable duration model and upsampling mechanism enables error gradients to be propagated through all operations in the network.
Experimental results show that Parallel Tacotron~2 can synthesize naturally sounding speech efficiently and outperforms baselines in subjective naturalness.
Its duration control capability is demonstrated.

The rest of the paper is organized as follows.
Section~\ref{sec:related_work} discusses the relationship between Parallel Tacotron~2 and prior work.
Section~\ref{sec:ptaco_basics} revisits the duration model of Parallel Tacotron and elaborates on the short-comings of the use of supervised duration signals. 
Section~\ref{sec:dur_model} introduces the architectural changes that enables the model to learn the mapping between the token and frame sequences as well as 
the new iterative reconstruction loss based on Soft-DTW. 
The paper is wrapped up with the experimental results in Section~\ref{sec:experiments} and some concluding remarks.

\section{Related Work}
\label{sec:related_work}

The proposed Parallel Tacotron~2 is a duration-based non-autoregressive neural TTS model which does not require supervised duration signals, like \cite{zeng-icassp-2020,lim-arxiv-2020,miao2020efficienttts,miao-icassp-2020,donahue-arxiv-2020}.

AlignTTS \cite{zeng-icassp-2020}, JDI-T \cite{lim-arxiv-2020}, and EfficientTTS \cite{miao2020efficienttts} train an alignment network jointly with a TTS model to produce alignments between tokens and frames. 
Durations are extracted from the alignments and then used as targets for the duration predictor in the TTS model.
To upsample the token sequence with durations, AlignTTS and JDI-T employ the length regulator \cite{ren2019fastspeech}, whereas EfficientTTS uses the Gaussian kernel mechanism  \cite{donahue-arxiv-2020}.
Differently, Parallel Tacotron~2 uses a combination of differentiable duration modeling and learned upsampling to extract alignments and model durations.

Instead of using token-level durations, Flow-TTS \cite{miao-icassp-2020} models the total number of frames in an utterance given tokens.
The token-to-frame mapping is computed using a dot-product attention with  frame-level queries represented by sinusoidal positional embeddings.
Since there is no token durations, fine-grained control of rhythm/pace of synthesized speech is difficult.
Although Parallel Tacotron~2 also uses the dot-product attention for upsampling, its attention matrix is derived from the predicted durations and the internal token-level representation.
Furthermore, Flow-TTS teacher-forces the total number of frames to enforce the lengths of target and prediction to be equal. Meanwhile, Parallel Tacotron~2 does not require it as the length mismatch is handled by Soft-DTW.
 
EATS \cite{donahue-arxiv-2020} relies on a block-alignment assumption, where any random fixed-length block is assumed to be located in the same location in both prediction and target spectrograms. 
Within the assumed aligned-blocks ($\sim$47 frames), Soft-DTW \cite{softdtw} is used as the reconstruction loss.
Although Parallel Tacotron~2 also uses Soft-DTW, it does not rely on such an assumption; Soft-DTW is performed over entire utterance.

\section{Duration Modeling in Parallel Tacotron}
\label{sec:ptaco_basics}

Parallel Tacotron \cite{ptaco} requires supervised token durations to be provided by an external aligner. 
Although these aligners usually provide reasonable alignments, they are not necessarily the best form to the decoder as they are not jointly trained. 
Parallel Tacotron relies on the length regulator \cite{ren2019fastspeech} to upsample encoder outputs according to the token durations.
The length regulator requires integral durations, \ie $d_i\in\mathbb{N}$ where $d_i$ denotes the duration of the $i$-th token.
Therefore, durations need to be rounded $\mathbb{R}\rightarrow\mathbb{N}$ before length regulation. 
This rounding introduce two problems.
First, it injects a rounding error. 
Although we can minimize the rounding error with a simple rounding algorithm, 
the error persists and needs to be dealt with the network.
Second, the rounding operation used in \cite{ptaco} is not differentiable, thus error gradient is not propagated through the operation.

Lastly, to use $L_1$ loss as a reconstruction loss, Parallel Tacotron needs to use teacher forcing \cite{teacherforcing} over the length regulator with the target durations. 
Without teacher forcing, the target and predicted spectrograms would not be of the same length. 
Teacher-forcing the durations can cause a discrepancy between training and inference; target durations are used at training time while predicted durations are used at inference time. 
Furthermore, when teacher forcing is used over durations, no error gradients are propagated from the reconstruction loss to the duration prediction.
This prevents the joint optimization of the duration predictor and the decoder so as to minimize the reconstruction loss.

\section{Parallel Tacotron~2}
\label{sec:dur_model}

\begin{figure}[t]
\centering
\includegraphics[width=0.35\textwidth]{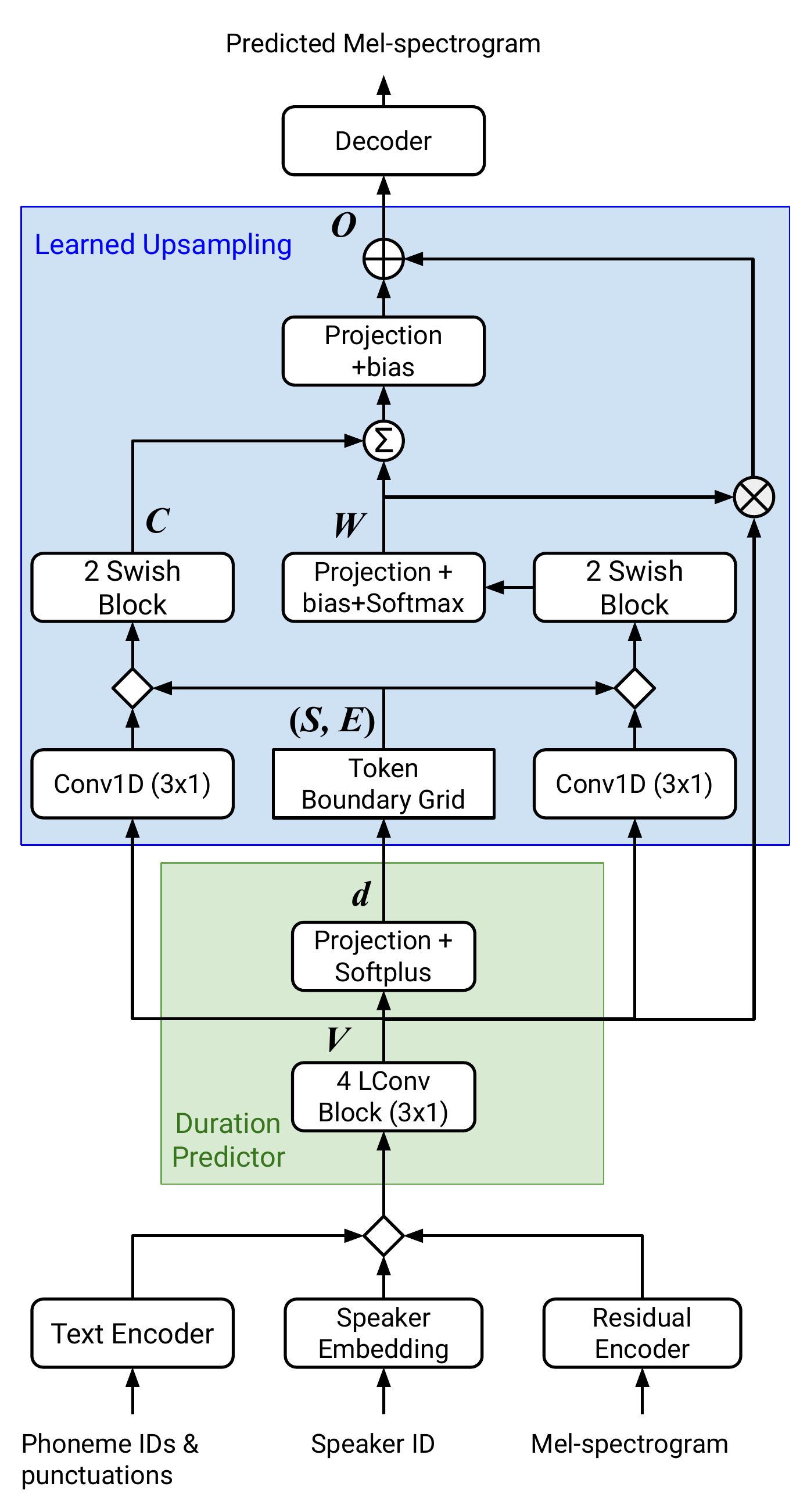}
\caption{Block diagram of the Parallel Tacotron~2 model. The green and blue blocks in the diagram describe the duration predictor and learned upsampling, respectively. $\Sigma$ refers to the \texttt{einsum('tk,tkp->tp', W, C)} operation.
Refer to the diagram in \cite{ptaco} for details of the decoder, text encoder, and residual encoder.}
\label{fig:architecture}
\end{figure}

This section introduces the proposed Parallel Tacotron~2 model, specifically differentiable duration modeling and learned upsampling, which don't require supervised duration signals.
The network architecture is illustrated in Fig.~\ref{fig:architecture}. 
It is designed to enable error gradients to be propagated through duration modeling, which is essential for automatically learning reasonable alignments between token and frame sequences without supervision.
The design includes (1) to propagate error gradients, durations $\bd$ and all operations are on real numbers $\mathbb{R}$ (as opposed to natural numbers $\mathbb{N}$);
(2) assumptions and discrepancy between training and inference such as teacher forcing are also eliminated to enable the network to learn the token-to-frame mapping.

\subsection{Differentiable Duration Modeling \& Upsampling}
The Duration Predictor of \cite{ptaco} is used to get a sequence of representation $\bV = \left\{ \bv_1, \dots, \bv_K \right\}$ and token durations $\bd = \{ d_1, \dots, d_K \}$,\footnote{The network  doesn't rely on the zero-length classification from \cite{ptaco}.} where $K$ denotes the number of tokens and $\bv_k$ is a $M \times 1$ column vector. 
Instead of relying on externally provided per-token supervised duration signals, here we define a duration loss $\mathcal{L}_{\mathrm{dur}}$ by using the total target frame duration $T$ only as
\begin{align}
    \mathcal{L}_{\text{dur}} = \frac{1}{K} \left\lVert T - \displaystyle \sum_{k=1}^K {d_k} \right\rVert_1.
\end{align}
To upsample $\bV = \{\bv_1,\dots,\bv_K\}$ into $\bO = \{\bo_1,\dots,\bo_T\}$, the learned upsampling module is designed to represent a function to map token durations to an attention matrix.

First, \emph{Token Boundaries} $[s_k,e_k)_{k=1}^K$ are computed from the token durations as
\begin{equation}
s_k=\sum^{k-1}_{i=1}{d_i}, \qquad e_k = s_k + d_k,
\end{equation}
Second, the token boundaries are mapped into two $T \times K$ grid matrices $\bS$ and $\bE$, which give distances to the boundaries of token $k$ at time $t$ as
\begin{equation}
S_{tk} = t-s_k, \qquad E_{tk} = e_k-t,
\end{equation}
where $S_{tk}$ and $E_{tk}$ are the $(t,k)$-th elements of $\bS$ and $\bm{E}$, respectively.
A $T \times K$ attention matrix $\bW$ is computed from $\bS$, $\bE$, and $\bV$ as
\begin{equation}
\label{equ:W}
\bW = \text{Softmax}\left( \text{MLP} \left( \bS, \bm{E}, \text{Conv1D}(\bV) \right) \right),
\end{equation}
where $\text{MLP}(\cdot)$ denotes a multi-layer perceptron-based learnable function.
A $T \times K \times P$ shaped \textit{Auxiliary Attention Context} tensor $\bC = \left[ \bC_1, \dots, \bC_P\right]$ is also learned in a similar way as
\begin{equation}
\label{equ:C}
\bC = \text{MLP} \left( \bS, \bm{E}, \text{Conv1D}(\bV) \right),
\end{equation}
where $\bC_p$ is a $T \times K$ matrix from $\bC$. 
$\bC$ can also be viewed as an auxiliary multi-headed attention-like information for the decoder.
We hypothesize that this extra representation smooths the optimization problem and helps SGD converge towards a good solution.
A preliminary investigation showed that this auxiliary representation helps the network train well.
Notably, there are no additional sinusoidal positional embeddings, meaning that this network learns a positional and contextual representation by itself.
The upsampled representation $\bO$ is computed as a regular attention matrix multiplication between $\bV$ and $\bW$ and a reduction 
of $\bC$ as
\begin{equation}
  \label{equ:O}
    \bO = \bW \bV 
    + \begin{bmatrix} (\bW \odot \bC_1) \bOne_K & \dots & (\bW \odot \bC_P) \bOne_K \end{bmatrix} \bA  
\end{equation}
where $\odot$ denotes element-wise multiplication, $\bOne_K$ is a $K \times 1$ column vector whose elements are all 1, and $\bA$ is a $P \times M$ projection matrix.\footnote{The operation among $\bW$, $\bC$, and $\bOne_K$ in Eq.~\eqref{equ:O} can be written by the \texttt{einsum} operation in a simple manner as \texttt{einsum('tk,tkp->tp', W, C)}, where \texttt{einsum} is a
multi-dimensional linear algebraic array operation in the Einstein summation convention. }

Both MLPs in Eqs.~\eqref{equ:W} and \eqref{equ:C} are modeled by two projection layers with  Swish activation \cite{swish} and bias. 
Both projection layers  in Eqs.~\eqref{equ:W} have output dimension of 16 and in \eqref{equ:C} the dimension is 2 (i.e. $P=2$). The MLP in Eqs.~\eqref{equ:W} has a third projection layer and bias with output dimension 1 which is fed to the Softmax activation function.
$\text{Conv1D}(\bV)$ has kernel-width 3, output dimension 8, batch normalization, and Swish activation.

\subsection{Reconstruction Loss using Soft-DTW}
\label{sec:loss}
Since the predicted and target spectrogram can have different lengths, the regular $L_1$ loss cannot be applied. 
To circumvent this mismatch a loss based on Soft-DTW \cite{softdtw} is used. Soft-DTW is a differentiable variant of the well known DTW dynamic programming (DP) with the following recursion:
\begin{displaymath}
r_{i, j} = \mathrm{min}^\gamma 
\begin{cases}
    r_{i-1, j} &+ \left\lVert \bx_{i-1}-\bar{\bx}_{j}\right\rVert_1 + \text{warp}\\
    r_{i, j-1} &+  \left\lVert \bx_{i}-\bar{\bx}_{j-1}\right\rVert_1 +\text{warp}\\
    r_{i-1, j-1} &+ \left\lVert \bx_{i-1}-\bar{\bx}_{j-1}\right\rVert_1 \\
\end{cases}
\end{displaymath}
where $r_{i,j}$ denotes the distance between target spectrogram frames from 1 to $i$ and predicted ones from 1 to $j$ with the best alignment, $\mathrm{min}^\gamma$ is a generalized minimum operation with a smoothing parameter $\gamma$, $\text{warp}$ is a warp penalty, and $\bx_i$ and $\bar{\bx}_j$ are the target and predicted spectrogram frames in time $i$ and $j$, respectively.

As in Parallel Tacotron, an iterative spectrogram loss \cite{dejavu} is used. 
Specifically, the decoder stack is based on 6 lightweight convolution (LConv) blocks \cite{wu2019pay}, 
with the output of each block being used to predict the output spectrogram. 
Similarly, Parallel Tacotron~2 iteratively predicts the output and Soft-DTW is used to compute the loss for each predicted output.

Note that the Soft-DTW setup here is computationally intensive. 
For the full dynamic programming all pairwise frame distances need to be computed with complexity $O(T^2)$. 
We implemented custom differentiable diagonal band operations 
where the width of diagonal band is fixed at 60.
The warp penalty
is set to $128$ and $\gamma=0.05$.

\subsection{Fine-grained Token Level VAE}
\label{sec:fine_vae}

The residual encoder in Parallel Tacotron~2 uses the same fine-grained token-level VAE as in Parallel Tacotron \cite{ptaco}, which consists of 5 LConv blocks. 
The Duration Predictor takes a posterior latent at training time and a zero vector (mean of prior) at inference time.
Note that the input to the residual encoder is conditioned on regular sinusoidal positional embeddings, rather than the supervised positional embeddings as in \cite{ptaco}. 

\subsection{Training Objective}
The overall loss function for Parallel Tacotron~2 with fine-grained VAE becomes
\begin{equation}
 \mathcal{L} = \frac{1}{LT} \sum_{l=1}^L{\mathcal{L}^{(l)}_{\text{spec}}} + \lambda_{\text{dur}} \mathcal{L}_{\text{dur}} + \beta D_\mathrm{KL}
 \end{equation}
where $\mathcal{L}_{\text{spec}}^{(l)}$ is the Soft-DTW $L_1$ spectrogram reconstruction loss for the $l$-th iteration in the spectrogram decoder, $\mathcal{L}_{\text{dur}}$ is the average duration $L_1$ loss, $D_\mathrm{KL}$ is the KL divergence between prior and posterior from the residual encoder.

\section{Experiments}
\label{sec:experiments}
\input{experiments.tex}

\section{Conclusions}
A non-autoregressive neural TTS model called Parallel Tacotron~2 was proposed. It outperforms the baseline supervised Parallel Tacotron in naturalness, preference tests, and with faster inference. The core invention of this work is based on a novel learned attention mechanism that learns the token-to-frame mapping without assumptions. Future work includes investigating better fine-grained variational models and relying on the second-order directional derivative of Soft-DTW to further guide the learnt alignment. Moreover, we intend to investigate whether Learned Upsampling and the Auxiliary Attention Context can help improve attention in other domains.

\clearpage


\bibliographystyle{IEEEtran}

\bibliography{main}

\end{document}

%% file: mymacro.tex
\bmdefine{\bO}{O}
\bmdefine{\bC}{C}
\bmdefine{\bc}{c}
\bmdefine{\bo}{o}
\bmdefine{\bW}{W}
\bmdefine{\bmu}{\mu}
\bmdefine{\bQ}{Q}
\bmdefine{\bq}{q}
\bmdefine{\bw}{w}
\bmdefine{\bU}{U}
\bmdefine{\bL}{L}
\bmdefine{\bu}{u}
\bmdefine{\bZero}{0}
\bmdefine{\bI}{I}
\bmdefine{\bR}{R}
\bmdefine{\bP}{P}
\bmdefine{\br}{r}
\bmdefine{\be}{e}
\bmdefine{\bE}{E}
\bmdefine{\bmm}{m}
\bmdefine{\bsigma}{\sigma}
\bmdefine{\bSigma}{\Sigma}
\bmdefine{\bOmega}{\Omega}
\bmdefine{\bomega}{\omega}
\bmdefine{\bS}{S}
\bmdefine{\bA}{A}
\bmdefine{\bC}{C}
\bmdefine{\bM}{M}
\bmdefine{\bg}{g}
\bmdefine{\bs}{s}
\bmdefine{\bpsi}{\psi}
\bmdefine{\bPsi}{\Psi}
\bmdefine{\bphi}{\phi}
\bmdefine{\bPhi}{\Phi}
\bmdefine{\bPi}{\Pi}
\bmdefine{\bpi}{\pi}
\bmdefine{\bLambda}{\Lambda}
\bmdefine{\blambda}{\lambda}
\bmdefine{\bB}{B}
\bmdefine{\bb}{b}
\bmdefine{\bl}{l}
\bmdefine{\bd}{d}
\bmdefine{\bD}{D}
\bmdefine{\bY}{Y}
\bmdefine{\bG}{G}
\bmdefine{\bp}{p}
\bmdefine{\bxi}{\xi}
\bmdefine{\bmeta}{\eta}
\bmdefine{\bzeta}{\zeta}
\bmdefine{\bk}{k}
\bmdefine{\bK}{K}
\bmdefine{\bF}{F}
\bmdefine{\bv}{v}
\bmdefine{\bX}{X}
\bmdefine{\bx}{x}
\bmdefine{\by}{y}
\bmdefine{\bz}{z}
\bmdefine{\bZ}{Z}
\bmdefine{\bcalX}{\mathcal{X}}
\bmdefine{\bH}{H}
\bmdefine{\bh}{h}
\bmdefine{\bcalH}{\mathcal{H}}
\bmdefine{\bV}{V}
\bmdefine{\bOne}{1}
\def\diag{\operatorname{diag}}
\def\idiag{\operatorname{diag}^{-1}}
\def\tr{\operatorname{tr}}
\def\vec{\operatorname{vec}}
\def\Gauss{\mathcal{N}}
\def\Qf{\mathcal{Q}}
\def\calM{\mathcal{M}}
\def\Ind{\mathrm{I}}
\def\Err{\mathcal{E}}
\def\Data{\mathcal{D}}
\def\Loss{\mathcal{L}}

\definecolor {GoogleRed}   {rgb}{0.97265625, 0.00390625, 0.00390625}
\definecolor {GoogleBlue}  {rgb}{0.0078125,  0.3984375,  0.78125}
\definecolor {GoogleYellow}{rgb}{0.9453125,  0.70703125, 0.05859375}
\definecolor {GoogleGreen} {rgb}{0.0,        0.57421875, 0.23046875}
\def\GoogleLogo{\textsf{\textcolor{GoogleBlue}{G}\textcolor{GoogleRed}{o}\textcolor{GoogleYellow}{o}\textcolor{GoogleBlue}{g}\textcolor{GoogleGreen}{l}\textcolor{GoogleRed}{e}}}
\def\GoogleAILogo{\GoogleLogo~\textsf{AI}}

\hyphenation{Libri-TTS}
\hyphenation{Libri-Speech}
\hyphenation{Wave-Net}
\hyphenation{Wave-RNN}
\hyphenation{Taco-tron}

\def\ie{\textit{i.e.,}}
\def\eg{\textit{e.g.,}}

%% file: experiments.tex
\subsection{Training Setup}
A proprietary speech dataset from \cite{ptaco} containing 405 hours of speech data (347,872 utterances) including 45 speakers in 3 English accents (31 US English speakers, 8 British English, and 5 Australian English speakers) was used.
The models from \cite{ptaco} were used as the baselines.\footnote{We'd like to underline that the Tactron~2 baseline used is likely of significantly better quality than other Tacotron~2 baselines used in the literature. Notably both the reduction factor, GMM attention, and training details are important for good quality.}

A proprietary text normalization engine was used to produce a phoneme sequences given input text.
We used phonemes and punctuation marks as input tokens.
Parallel Tacotron~2 models were trained using the Adam optimizer with the learning schedule from \cite{transformer} with 10k warmup steps. 
The model was trained for 500k steps with a batch size of 2,048 using Google Cloud TPUs. 
For the fine-grained token-level VAE, a KL-weight schedule was used where $\beta$ was increased linearly to 1.0 from step 6K to 50K and $\lambda_{dur}=100$.
Both baseline and proposed models were combined with the same pretrained WaveRNN neural vocoder \cite{kalchbrenner2018efficient} to reconstruct audio signals from predicted mel-spectrograms.

\subsection{Evaluation Setup}
We perform two sets of experiments comparing Parallel Tacotron~2 to the baselines from \cite{ptaco} in several diverse multi speaker evaluations on an internal evaluation platform. The sentences were different from the training data and used in  previous papers.
They were synthesized using 10 US English speakers (5 male \& 5 female) in a round-robin style.
The amount of training data for the evaluated speakers varied from 3 hours to 47 hours. 

In the first experiment, we used the same evaluation set as in \cite{ptaco}, and conducted subjective evaluations over 1,000 sentences. We further performed several additional direct comparisons using the diverse evaluation sets from \cite{iris}. In addition, we performed direct comparisons using 1,000 sentences from the same hold-out set as in \cite{ptaco}. Finally, we conduct  direct comparisons between our models and natural speech using the hold-out set.

Naturalness was evaluated through subjective listening tests, including 5-scale Mean Opinion Score (MOS) tests and side-by-side preference tests. 
For the MOS tests, a five-point Likert scale score (1: Bad, 2: Poor, 3: Fair, 4: Good, 5: Excellent) was adopted with rating increments of 0.5.
For the preference tests each rater listened to two samples then rated each with integral scores $[-3,3]$; where a positive score indicated that the first sample sounded better than the second one \cite{transfer, tacotron2}.
Raters were instructed to use headphones.

Although the baseline models are the same as in \cite{ptaco}, all experimental evaluations have been performed anew. We further normalized loudness to ensure that the average loudness of each model was the same; this reduces bias in the evaluation results.

\subsection{Experimental Results}

The first experiment Table~\ref{unsup_vs_baselines} evaluated the proposed model against the baselines on the evaluation set from \cite{ptaco}. As can be seen, Parallel Tacotron~2 outperforms the baselines in the preference tests.

\begin{table}[ht]
\caption{Subjective MOS and preference scores between Parallel Tacotron~2 and baseline models from \cite{ptaco}. Positive preference scores indicate that  Parallel Tacotron~2 was rated better than the baseline.}
\begin{tabular}{lrr}
\toprule  
\textbf{Model} & \textbf{MOS} & \textbf{Preference}
\\\midrule
Parallel Tacotron~2	& $4.40 \pm 0.05$ & Reference
\\\midrule
Tacotron~2 \cite{tacotron2} &	$4.40 \pm 0.05$ &	$\bf{0.19 \pm 0.09}$\\\midrule
Parallel Tacotron\cite{ptaco}\\
\quad Global VAE &	$4.37 \pm 0.05$ &	$\bf{0.14 \pm 0.08}$\\
\quad Fine VAE &	$4.40 \pm 0.05$& $0.03 \pm 0.08$ \\
\bottomrule
\end{tabular}
\label{unsup_vs_baselines}
\end{table}

The second series of experiments Table~\ref{evaluation_sxs} performed additional preference tests on four other evaluation sets. Specifically, three evaluation sets from \cite{iris} plus 1,000 utterances from the hold-out set.  Parallel Tacotron~2 outperformed the baselines in the Rapid evaluation set, and matched the baselines in other evaluations.
We found that prosody of synthesized speech in this evaluation set sounded significantly more natural than the baseline.

\begin{table}[ht]
\caption{Subjective preference scores between Parallel Tacotron~2 and the baseline models. 
The ``Rapid'' evaluation set consisted of 100 questions prompting the user to rephrase, \eg ``Sorry, what was that?''
The ``Questions'', and ``Hard'' evaluation sets correspond to those 
 used in \cite{iris}. Positive preference scores indicate that  Parallel Tacotron~2 was rated better than the baseline.}
\begin{tabular}{l|l|l|l}
\toprule  
& \multicolumn{3}{c}{\textbf{Baselines}} \\
&  Tacotron~2 \cite{tacotron2} &  \multicolumn{2}{|c}{Parallel Tacotron \cite{ptaco}} \\
\textbf{Eval. set}  &  w/o VAE                     & Global VAE & Fine VAE \\
\midrule
Rapid     &	$\bf{0.69 \pm 0.16}$ & $\bf{0.74 \pm 0.13}$  & $\bf{0.38 \pm 0.14}$ \\
Questions &	$0.07 \pm 0.12$ & $0.02 \pm 0.10$  & $0.08 \pm 0.10$ \\
Hard      & $0.03 \pm 0.11$ & $\bf{0.19 \pm 0.10}$ & $0.04 \pm 0.09$ \\
Hold-out  & $0.02 \pm 0.08 $ & $0.00 \pm 0.08$ & $0.03 \pm 0.07$ \\
\bottomrule
\end{tabular}
\label{evaluation_sxs}
\end{table}

The third experiment Table~\ref{preference_human} compared Parallel Tacotron~2 and natural speech on the 1,000 utterances from the hold-out set. 
It can be seen from the table that the neural TTS models are doing well compared to human speech. Parallel Tacotron~2 is even rated better than Natural speech. However, we underline that our training data and the hold-out set contain artifacts which raters identify. As such, there is still a room for further improvement.

\begin{table}[ht]
\caption{Subjective MOS and preference scores (between natural speech from the hold-out test set and Parallel Tacotron~2). Positive preference scores indicate that synthesized speech was rated better than Natural speech.}
\begin{tabular}{l|lr}
\toprule  
\textbf{Model} & \textbf{MOS} & \textbf{Preference}
\\\midrule
Natural speech	& $4.49 \pm 0.05$ & Reference \\	
\midrule
Parallel Tacotron & \\
\quad Global VAE		& & $-0.04 \pm 0.09$ \\
\quad Fine VAE	& $4.42 \pm 0.05$ & $-0.08 \pm 0.09$ \\
\midrule
Parallel Tacotron~2 &	$4.46 \pm 0.05$ &	$0.01 \pm 0.09$ 
\\\bottomrule
\end{tabular}
\label{preference_human}
\end{table}


\begin{figure}[t]
\centering
\includegraphics[scale=0.4]{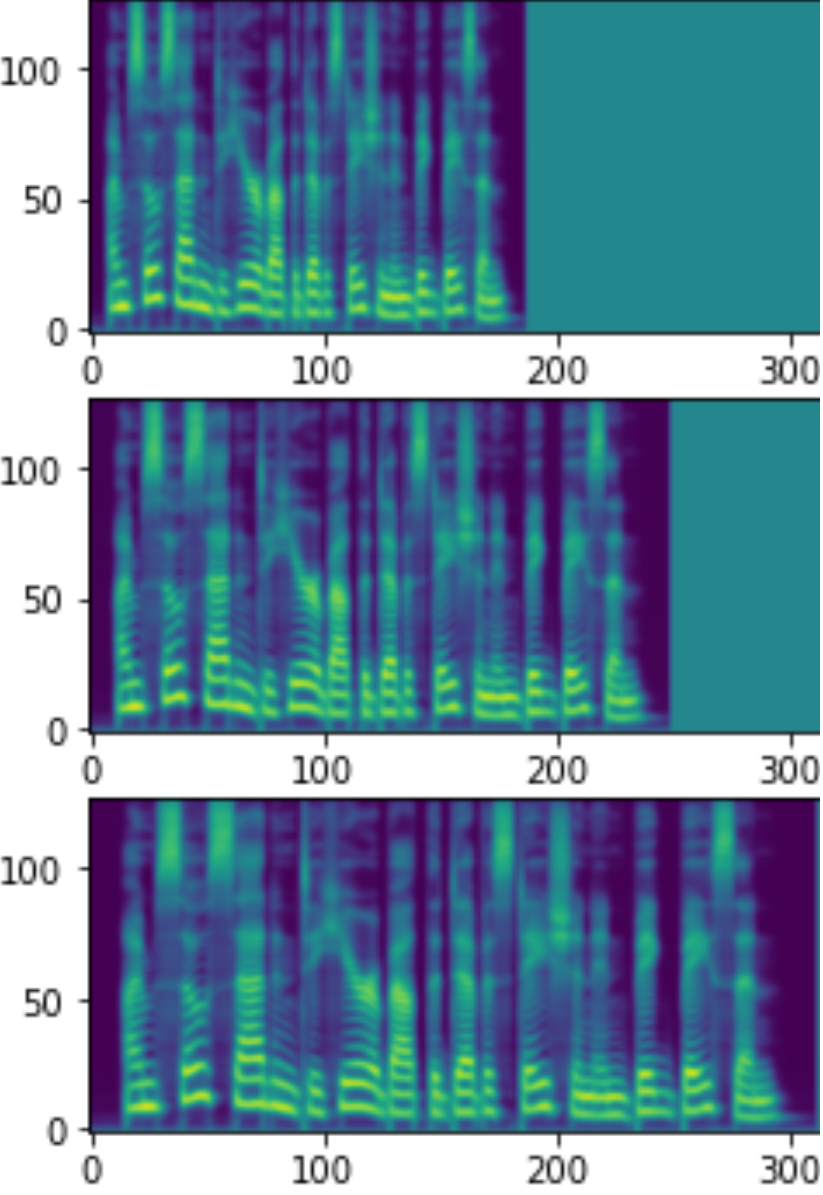}
\includegraphics[scale=0.4]{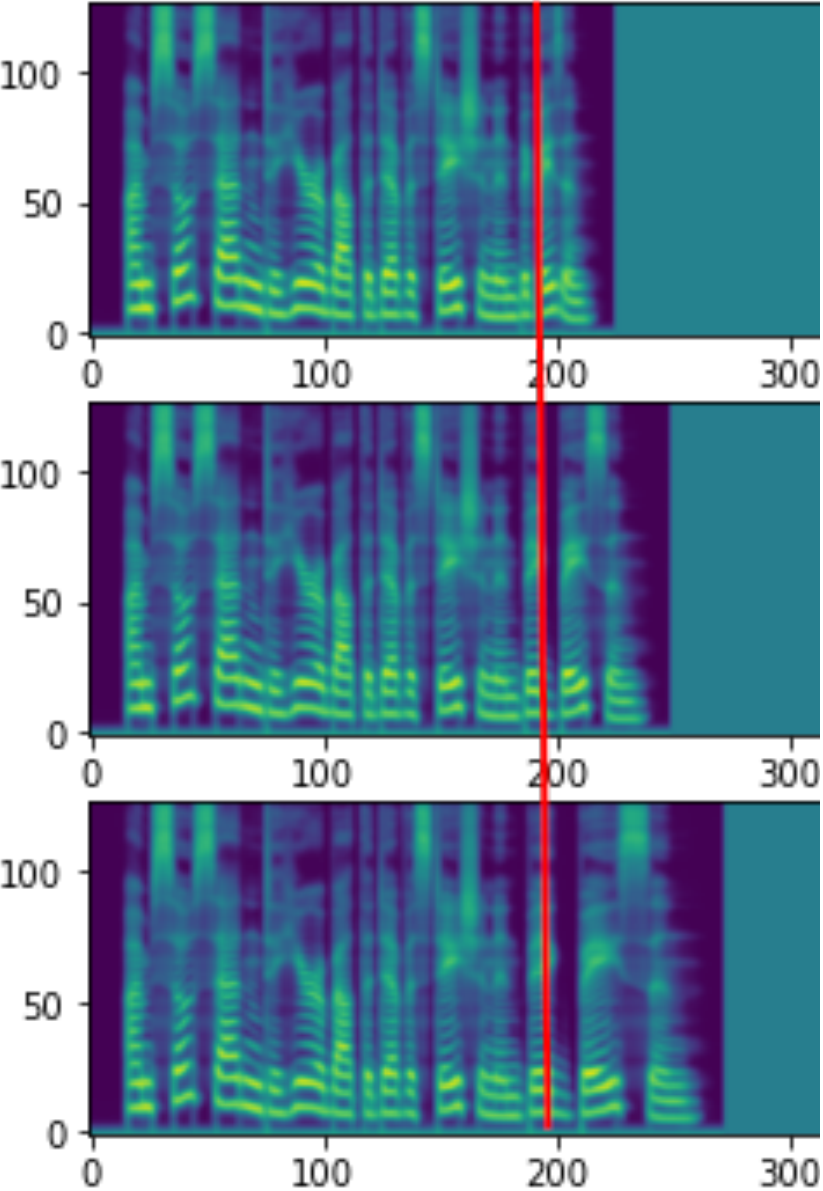}
\caption{Duration control of the sentence "I'm so saddened to hear about the devastation in Big Basin.". Left: Global pace control, all predicted durations are scaled by factors $[ 0.75, 1.0, 1.25 ]$.
Right: Word-level duration control, the ending "Big Basin" is scaled by $[0.5, 1.0, 1.5]$ }
\label{fig:pacecontrol}
\end{figure}

\subsection{Manual Control of Durations}
Since Parallel Tacotron~2 learns alignments between token and frames without duration supervision, it bears to question whether the learned alignments match actual token boundaries.  
Specifically, we still desire a model in which pace and duration are controllable.

The left part of Fig.~\ref{fig:pacecontrol} demonstrates that the total duration of the synthesized speech was controllable by scaling all predicted durations by a fixed factor. Likewise, the right part of Fig.~\ref{fig:pacecontrol} shows that the duration of individual words can also be controlled by scaling durations of individual tokens.